\documentclass{mem}
\usepackage{natbib}\usepackage{txfonts}\usepackage{balance}
\usepackage{graphicx}
\usepackage[a4paper]{hyperref}
\idline{00}{001}
\begin{document}
\def\teff{$T\rm_{eff }$}
\def\kms{$\mathrm {km s}^{-1}$}

\title{
The Flaring Corona of UX\,Arietis
}

   \subtitle{}

\author{
E. \, Ros
\and M. \, Massi
          }

  \offprints{E. Ros}

\institute{
Max-Planck-Institut f\"ur Radioastronomie --
Auf dem H\"ugel 69, D-53121 Bonn, Germany
\email{ros@mpifr-bonn.mpg.de}
}

\authorrunning{Ros \& Massi}

\titlerunning{The Flaring Corona of UX\,Ari}

\abstract{
Here we present observational results on
the RS\,CVn star UX\,Arietis from four very-long-baseline
interferometry observations
distributed in time to cover the rotational period of 6.44\,days. 
The data are better fit by two Gaussian
components than by the usual core-plus-halo model. 
In the first three days the sizes of the
two components did not change much from hour to
hour but their relative
position and orientation changed from day to day. The origin of
this evolution can be explained by geometrical factors
(i.e., star rotation). The fourth day a large flare
occurred and dramatic changes in the sizes of the
Gaussian components were seen.
}
\maketitle{}

RS\,CVn stars are binary systems characterized by an intense coronal activity
at X-ray, UV and radio wavelengths. One of the most active sources at radio
wavelengths is the system UX\,Arietis, formed by a G5V and a K0\,IV star,
with an orbital period of 6.44 days and an orbit diameter
of 1.72\,mas \citep{Lestrade99}. 
Direct evidence of large structures with sizes comparable to the
binary system have been provided by VLBI observations 
(\cite{Mutel85},
\cite{Massi88},
\cite{Beasley96},
\cite{Franciosini99}).  The latter, made at a post-flare phase,
show a clear variation of the source
morphology from day to day.  
The morphology can be 
reproduced with a model consisting of two extended elliptical
Gaussian components 
with slowly decaying flux densities, that 
changed their relative position from day to day.
These are interpreted to be two radio loops emerging from the
position of the optical spots (see \cite{Elias95}), 
intruding into each other.

\begin{figure*}[tb]
\begin{center}
{\includegraphics[clip=true,angle=-90,width=0.95\textwidth]{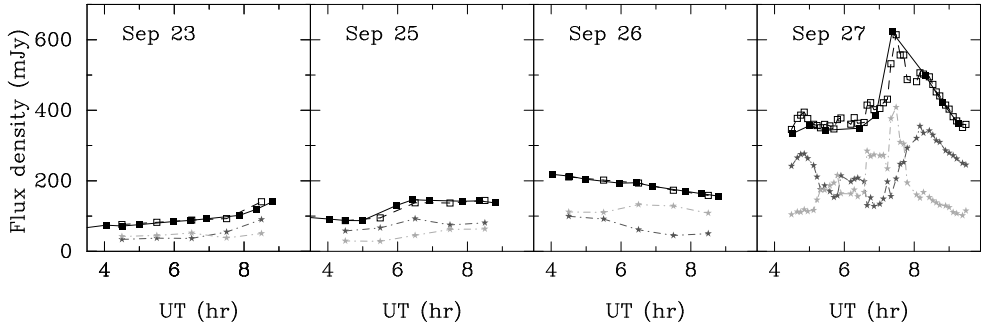}}
\end{center}
\caption{\footnotesize
Flux densities during the four observing runs. 
Filled boxes represent the single-dish Effelsberg
measurements \cite{Massi05}. Empty boxes show the 
total flux densities obtained from
model fitting the visibility data with the a priori calibration from 
the observing logs (uncertainties 5\,\%).  Stars denote flux densities 
for the two model fit components (A in dark grey, B in light grey).
}
\label{fig:ros_flux}
\end{figure*}
\begin{figure*}[p]
\begin{center}
{\includegraphics[clip=true,width=0.83\textwidth]{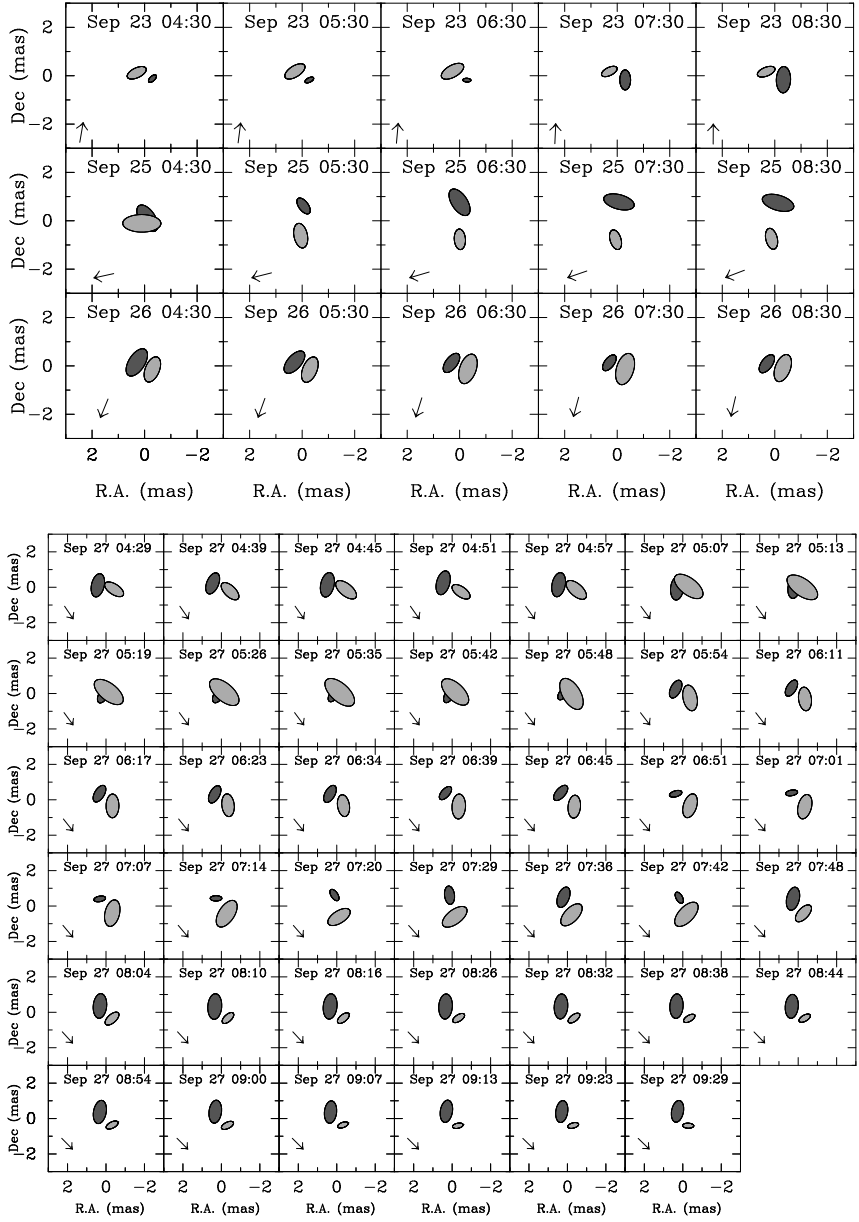}}
\end{center}
\caption{\footnotesize
{\bf Top:}
Relative positions and sizes for the elliptical Gaussians
that model the visibilities of UX\,Arietis at epochs 23, 25, and 
26 september 2001. We used 1\,hr segments to perform the model fitting.  
The component A is shown with dark gray tones, the 
component B with light gray.  
The arrows at the bottom, left represent the P.A. from the optical
phase (not projected at the orbit inclination of 60\degr).  A
phase 0\degr implies the K0\,IV star to be to the south, and between the G5V
star and the observer.
The axis ratio
of both Gaussian functions was fixed at a value of 0.5.
We solved for flux density,
position, major axis,
and position angle for the major axis.  
{\bf Bottom}
Relative positions and sizes of A and B for Sep 27.  The
time binning is made scan by scan of length 4\,min.
}
\label{fig:ros_beams}
\end{figure*}

\begin{table}[htb]
\caption{Observation log}
\label{table:log}
\begin{center}
\vspace*{-12pt}
\begin{tabular}{@{}lllll@{}}
\hline
Exp.$^*$ & Date$^\star$ & Orb.\ Phase \\
\hline
{\tt BM140B} & 2001 Sep 23 & 0.46 to 0.49 \\
{\tt BM140C} & 2001 Sep 25 & 0.78 to 0.81 \\
{\tt BM140E}$^\dag$ & 2001 Sep 26 & 0.93 to 0.97 \\
{\tt BM140D} & 2001 Sep 27 & 0.10 to 0.13 \\
\hline
\end{tabular}
\vspace*{-8pt}
\end{center}
\scriptsize{
$^*$ The array used was the VLBA+Effelsberg in all cases.
$^\star$ The UT range was 03:30 to 09:30 for all four observing runs.
$^\dag$ Observation initially scheduled on Sep 21,  but 
shifted due to scheduling problems in Effelsberg.}
\end{table}

We performed VLBI observations (see Table~\ref{table:log}) 
with the Very Long Baseline Array (VLBA) and the 100-m Effelsberg telescope
during a complete rotation cycle of the system
to track the motion of the stellar radio-sphere.
We also monitored the total and polarized flux density simultaneously with
Effelsberg to distinguish
temporal flux density variations from structural variations
(see Fig.~\ref{fig:ros_flux}).
Activity in UX\,Arietis was present during all epochs, and a flaring event 
occurred during the last observing
day, Sep 27.  

\begin{figure}[tbh]
\begin{center}
{\includegraphics[clip=true,width=0.45\textwidth]{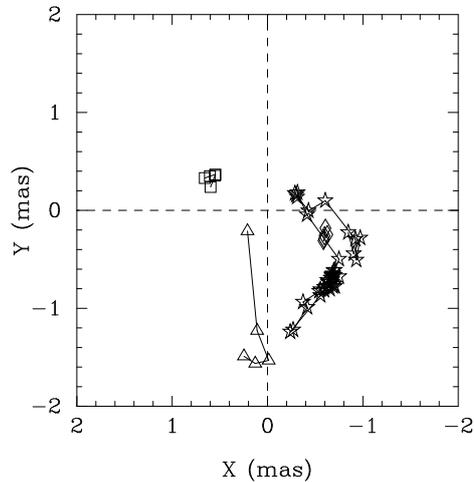}}
\end{center}
\caption{\footnotesize
Relative positions in the sky of component B with respect to component A
for the four observing epochs 
(squares: Sep 23; 
triangles: Sep 25;
diamonds: Sep 26;
stars: Sep 27).
}
\label{fig:ros_xy}
\end{figure}

Since UX\,Arietis changed in flux density during the observations,
we could not image the radio source using the whole data set,
since the assumption of an unchanging source, upon which the
aperture synthesis principle is based, was violated.
We analyzed the data by splitting them in time in 1\,hr segments for
the first three epochs, and for each observing scan of length 4\,min 
the fourth epoch.
The results of this procedure are shown in Fig.~\ref{fig:ros_beams}
and Fig.~\ref{fig:ros_xy}.

We interpret our observing results as follows:\\[2pt]
{\bf 1st day; $\langle$$\varphi$$\rangle$=0.5:} The spotted
hemisphere is eclipsed by the K0 IV
star itself -–- the optical spots were not
visible except for those at high latitudes
(60\degr)\\[2pt]
{\bf 2nd day; $\langle$$\varphi$$\rangle$=0.8:} the distance between
components was maximized, the K0 IV
star was to the SW of the GV5 star.\\[2pt]
{\bf 3rd day; $\langle$$\varphi$$\rangle$=0.9:} the K0 IV star was in
the foreground\\[2pt]
{\bf 4th day; $\langle$$\varphi$$\rangle$=0.1:} The big flare –--
observations revealed dramatic changes
in flux density and size of the model-fit
components.

\begin{acknowledgements}
The VLBA is operated by the National Radio Astronomy Observatory 
as a facility of the National Science Foundation 
operated under cooperative agreement by Associated Universities, Inc.  
The results presented here are based on observations with the 
100-m telescope of the MPIfR (Max-Planck-Institut für Radioastronomie) 
at Effelsberg.
\end{acknowledgements}

\bibliographystyle{aa}

\end{document}